\documentclass[english,aps]{revtex4}
\usepackage{mathptmx}

\usepackage[T1]{fontenc}
\usepackage[latin9]{inputenc}
\usepackage{amssymb}

\makeatletter
\@ifundefined{textcolor}{}
{%
 \definecolor{BLACK}{gray}{0}
 \definecolor{WHITE}{gray}{1}
 \definecolor{RED}{rgb}{1,0,0}
 \definecolor{GREEN}{rgb}{0,1,0}
 \definecolor{BLUE}{rgb}{0,0,1}
 \definecolor{CYAN}{cmyk}{1,0,0,0}
 \definecolor{MAGENTA}{cmyk}{0,1,0,0}
 \definecolor{YELLOW}{cmyk}{0,0,1,0}
 }

\makeatother

\usepackage{babel}

\begin{document}

\title{Comment on ``Thermal fluctuations of magnetic nanoparticles'', cond-mat/arXiv:1209.0298}
\author{J.-L. D\'ejardin, H. Kachkachi, and J.-M. Martinez}
\address{PROMES CNRS UPR8521, Universit\'e de Perpignan Via Domitia, 52 avenue
Paul Alduy, 66860 Perpignan, France}
\begin{abstract}
We comment on some misleading and biased statements appearing in the manuscript ``Thermal fluctuations of magnetic nanoparticles'', 
cond-mat/arXiv:1209.0298, about the use of the damped Landau-Lifshitz equation in conjunction with the kinetic Langer theory for the 
calculation of the relaxation rate of magnetic nanoclusters. We provide simple scientific arguments, part of which is well known to 
the whole community, demonstrating that the authors overstate the issue and contradict a work they have co-published earlier. 

\end{abstract}

\maketitle
Here we would like to draw the reader's attention to the following
scientifically misleading and biased statement on page 46 of
the manuscript cond-mat/arXiv:1209.0298 : {}``..... \emph{Unfortunately,
some authors (see, e.g., Ref. 137 and 138) have ignored this property
of the Landau-Lifshitz equation and, in consequence, have used this
intrinsically under-damped equation in conjunction with the intrinsically
IHD Langer formula for the calculation of the escape rate in all damping
ranges. Thus the ensuing escape rate formulas {[}Refs. 137, 138{]}
are misleading and not valid for experimental comparison both at low
damping, where they coincide with the TST rate, and also in the IHD
range, $\alpha\gtrsim1$, where they predict nonphysical behavior of
the rate, namely, a rate in excess of the TST one.}'' In the sequel,
this review paper will be referred to as CK.

Let us now give a brief account of our scientific point of view concerning
the issue of damping in the Landau-Lifshitz equation and its use in
\cite{kac03epl,kac04jml,francoetal11prb}, Refs. (137, 138), together
with Langer's approach. This is, of course, known to the whole community
working in this area, and we apologize for having to reiterate it
once again %
\footnote{We deem it our duty to inform the reader that the second author has
already sent to Phys. Rev. B a comment on the article \cite{francoetal11prb},
Ref. 138 in CK. After we sent our reply and after one more round in
the referral process of Phys. Rev. B, we decided that the various
comments and replies be sent to a third referee. To the best of our
knowledge, Kalmykov's comment has not appeared in Phys. Rev. B.%
}.

It goes without saying that the damping issue is a subtle one and
takes on special relevance in magnetism. The issue dates back to 1935
when Landau and Lifshitz published their seminal paper on magnetics,
and has been since a strong point of debate and controversy through
hundreds of publications and conference proceedings, especially after
Gilbert proposed, in the 1955 MMM conference proceedings, a new form
for the magnetization damping. Despite this long period of investigation,
the issue has not been settled yet and the very origin of damping
eludes any simple interpretation. The main reason is that damping
is rooted in various kinds of correlation processes, both intrinsic
and extrinsic, which cannot be captured by a few phenomenological
parameters added to the equation of motion. It would be too long,
if not impossible, to give a fair account of the divers approaches
and interpretations of damping in magnetic systems. A brief account
can be found in Ref. \onlinecite{saslow09jap} {[}\emph{Landau-Lifshitz
or Gilbert damping ? That is the question}{]} and Ref. \onlinecite{hicmoo09prl}
{[}\emph{Origin of intrinsic Gilbert damping}{]}.

The equation of motion describing the magnetization dynamics with
a phenomenological damping parameter can be represented as one of
the following two well-known forms:
\begin{enumerate}
\item The Landau-Lifshitz equation (LLE) \begin{equation}
\frac{d\vec{M}}{dt}=-\gamma\vec{M}\times\vec{H}-\lambda\frac{\gamma}{M}\vec{M}\times\left(\vec{M}\times\vec{H}\right),\label{eq:LLE}\end{equation}
with $\lambda$ being a (dimensionless) dissipation parameter and
$\vec{H}$ the effective field.
\item The Landau-Lifshitz-Gilbert equation (LLGE)
\end{enumerate}
\begin{equation}
\frac{d\vec{M}}{dt}=-\gamma\vec{M}\times\vec{H}+\frac{\alpha}{M}\vec{M}\times\frac{d\vec{M}}{dt}.\label{eq:LLGE}\end{equation}
where $\alpha$ is another (dimensionless) dissipation parameter. 

Mathematically, the two equations (\ref{eq:LLE}) and (\ref{eq:LLGE})
are equivalent. Indeed, substituting for $d\vec{M}/dt$ on the right-hand
side of Eq. (\ref{eq:LLGE}) the same right-hand side and working
out the resulting double cross product $\vec{M}\times\left(\vec{M}\times\frac{d\vec{M}}{dt}\right)$,
and using $\vec{M}\cdot\frac{d\vec{M}}{dt}=0$ (because the module
of $\vec{M}$ is constant), we obtain\begin{equation}
\frac{d\vec{M}}{dt}=-\left(\frac{\gamma}{1+\alpha^{2}}\right)\vec{M}\times\vec{H}-\left(\frac{\alpha}{1+\alpha^{2}}\right)\frac{\gamma}{M}\vec{M}\times\left(\vec{M}\times\vec{H}\right)\label{eq:LLGE2LLE}\end{equation}
which is just Eq. (\ref{eq:LLE}) upon making the following substitutions
\begin{equation}
\frac{\gamma}{1+\alpha^{2}}\rightarrow\gamma,\qquad\frac{\alpha}{1+\alpha^{2}}\rightarrow\lambda\label{eq:LLE2LLGE}\end{equation}
in the first and second terms, respectively. Note that this transformation
depends on the normalization used in both equations {[}see e.g. http://en.wikipedia.org/wiki/Landau-Lifshitz-Gilbert\_equation{]}.
Another useful form of Eq. (\ref{eq:LLGE2LLE}) consists in rewriting it in terms of the re-scaled time $\tau=t/\left(1+\alpha^{2}\right)$, 
and thus also re-scaling the N\'eel free diffusion time $\tau_\mathrm{N}$ \cite{garpal00acp}.

Further discussion of the two equations and their comparison can be
found in the textbooks \cite{gurmel96crcpress,stosie06springer}.
It is worth mentioning the work in Ref. \cite{laknak84prl} where
it is rigorously shown that the LLGE damping term is a mere re-scaling
of time by a complex constant. Moreover, it can be easily shown \cite{garpal00acp}
that the Fokker-Planck equations associated with the stochastic analogs
of the two equations (\ref{eq:LLE}, \ref{eq:LLGE}) are also identical. 

From the experimental point of view, there is no clear cut proof as
to which equation has to be used in general. In practice, based on
many investigations, it has been agreed upon that for small damping,
LLE and LLGE are almost the same and thereby the former is then assumed
to be more suited to small damping regimes. Indeed, for small damping,
the transformation in Eq. (\ref{eq:LLE2LLGE}) boils down to identity.
However, many workers obtain the LLE damping for low frequency, long
wavelength dynamics \cite{saslow09jap}. For high damping one would
expect a damping-dependent gyromagnetic ratio, but this effect has
still to be confirmed by experiments.

Now, the work \cite{francoetal11prb} (Ref. 137 in CK) uses the LLE
for obtaining the attempt frequency that enters the prefactor of Langer's
relaxation rate. Had we used the Landau-Lifshitz-Gilbert damping instead
we would have obtained expressions that can be recovered by making
the substitution (\ref{eq:LLE2LLGE}). A concrete example illustrating
this procedure is provided by the work published in Ref. \onlinecite{titovetal05prb}.
In this reference, Appendix B summarizes the analytical expressions
obtained in Ref. \onlinecite{kac03epl} ($1^{{\rm st}}$ paper in
Ref. 137 in CK), for a system of two exchange-coupled magnetic moments
using Langer's approach. Then, these analytical expressions were compared
in Figs. 2 and 3 of Ref. \onlinecite{titovetal05prb} to the results
of the fully independent numerical method of matrix continued fractions,
with a fairly good agreement. \smallskip{}

Last but not least, let us mention a few points about this review
that deserve special attention from the reader.
\begin{itemize}
\item It is curious how the authors' select their references when they write
{}``\emph{...some authors (see, e.g., Ref. 137 and 138) have ignored
this property of the Landau-Lifshitz equation''; }this is a rather
biased and non objective manner in reviewing the literature, at variance
with what a reader expects from a review article. Indeed, one\emph{
}of the well-known specialist in this area, Dmitry Garanin, and who
is acknowledged by the authors for his {}``\emph{direct or indirect}''
contribution to this review, has published fundamental and well-known
contributions with strong impact on the developments in this area
of physics. The authors seem to ignore the fact that all of Garanin's
papers exclusively use the Landau-Lifshitz equation. The reason is,
of course, scientifically motivated and is as explained above. In
the work \cite{garaninetal99pre} a kind of phase diagram was obtained
for uniaxial anisotropy with precise crossovers between various damping
regimes.
\item The authors claim that the analytical expressions published in Refs\emph{.
}137, 138 are misleading and {}``\emph{not valid for experimental
comparison}''. It is indeed very important to show some care for
comparison between theory and experiments. However, this manuscript
which is a big review that covers at least two decades fails to provide
a single comparison of experiments with any of the authors' own theoretical
work that goes beyond the N\'eel-Brown model. The only two figures 8
and 22 that show a comparison between experiments and N\'eel-Brown model,
are borrowed from the literature. A successful comparison between
the N\'eel-Brown model and experimental measurements on single magnetic
nanoparticles was achieved many years ago by W. Wernsdorfer et al. in the
seminal work \cite{weretal97prl,wernsdorferetal97etal,jametetal01prl}. 
\item The $2^{{\rm nd}}$ article in Ref. 137 in CK (which is Ref. \onlinecite{kac04jml})
was published as a review article in the special edition of the Journal
of Molecular Liquids that was edited and prefaced by the first author
of the review CK. This article summarized the main steps of Langer's
calculation of the relaxation rate \cite{lan68prl,lan69ap} and clearly
started the validity of the approach with respect to damping.
\end{itemize}
It is regretful that this big review does not provide a wider and more objective view
of the work available in the literature on the dynamics of magnetic
nanoclusters, for the benefit of a new comer to the field. It is also
unfortunate that the authors have not provided a discussion of the
huge amount of experimental work that shows the state-of-the-art understanding
of the real situation about these systems.

%\bibliography{hkbib}

\end{document}